\documentclass[traditabstract,letter]{aa} % for the letters 
\usepackage[hyperindex=true,colorlinks=true,citecolor=blue,linkcolor=blue,breaklinks=true]{hyperref}
\usepackage{amsmath}
\usepackage{graphicx}
\usepackage{txfonts}
\usepackage{natbib,twoopt}
\usepackage{color}

\bibpunct{(}{)}{;}{a}{}{,} %% natbib format like A&A and ApJ 
\newcommandtwoopt{\citeads}[3][][]{\href{http://adsabs.harvard.edu/abs/#3}% 
{\citealp[#1][#2]{#3}}} 
\newcommandtwoopt{\citepads}[3][][]{\href{http://adsabs.harvard.edu/abs/#3}% 
{\citep[#1][#2]{#3}}} 
\newcommandtwoopt{\citetads}[3][][]{\href{http://adsabs.harvard.edu/abs/#3}% 
{\citet[#1][#2]{#3}}}
\newcommandtwoopt{\citeyearads}[3][][]% 
{\href{http://adsabs.harvard.edu/abs/#3}{\citeyear[#1][#2]{#3}}}

\begin{document}

	\title{Multiplicity of Galactic Cepheids from long-baseline interferometry}
	\titlerunning{Multiplicity of Galactic Cepheids from long-baseline interferometry}

   	\subtitle{II.~The Companion of AX~Circini revealed with VLTI/PIONIER\thanks{Based on observations made with ESO telescopes at Paranal observatory under program ID 090.D-0010}}

	\author{ A.~Gallenne\inst{1},
      				A.~M\'erand\inst{2},
      				P.~Kervella\inst{3},
      				J.~Breitfelder\inst{2,3},
      				J.-B.~Le~Bouquin\inst{4},
    				J.~D.~Monnier\inst{5},
%				S.~Kraus\inst{5},
%				G.~H.~Schaefer\inst{6},  
					W.~Gieren\inst{1},
					B.~Pilecki\inst{1,6},
					\and G.~Pietrzy\'nski\inst{1,6}
%				L.~Szabados\inst{7}			
  				}
  				
  	\authorrunning{A. Gallenne et al.}

\institute{Universidad de Concepci\'on, Departamento de Astronom\'ia, Casilla 160-C, Concepci\'on, Chile
	\and European Southern Observatory, Alonso de C\'ordova 3107, Casilla 19001, Santiago 19, Chile
	\and LESIA, Observatoire de Paris, CNRS UMR 8109, UPMC, Universit\'e Paris Diderot, 5 Place Jules Janssen, F-92195 Meudon, France
	\and UJF-Grenoble 1/CNRS-INSU, Institut de Plan\'etologie et d’Astrophysique de Grenoble (IPAG) UMR 5274, Grenoble, France
	\and Astronomy Department, University of Michigan, 1034 Dennison Bldg, Ann Arbor, MI 48109-1090, USA
%	\and Harvard-Smithsonian Center for Astrophysics, 60 Garden Street, MS-78, Cambridge, MA 02138, USA
%  	\and The CHARA Array of Georgia State University, Mount Wilson CA 91023, USA
%  	\and Konkoly Observatory, Research Centre for Astronomy and Earth Sciences, Hungarian Academy of Sciences, H-1121 Budapest, Konkoly Thege Mikl\'os \'ut 15-17, Hungary
%  	\and School of Physics and Astronomy, University of St Andrews, North Haugh, St Andrews, Fife, KY16 9SS, UK
  	\and Warsaw University Observatory, Al. Ujazdowskie 4, 00-478, Warsaw, Poland}%\\
  
  \offprints{A. Gallenne} \mail{agallenne@astro-udec.cl}

   %\date{Received January 11, 2013; accepted February 6, 2013}

% \abstract{}{}{}{}{} 
% 5 {} token are mandatory
 
  \abstract
  % context heading (optional)
  % {} leave it empty if necessary  
   {}
  % aims heading (mandatory)
   {We aim at detecting and characterizing the main-sequence companion of the Cepheid AX~Cir ($P_\mathrm{orb} \sim $ 18\,yrs). The long-term objective is to estimate the mass of both components and the distance to the system.}
  % methods heading (mandatory)
   {We used the PIONIER combiner at the VLT Interferometer to obtain the first interferometric measurements of the short-period Cepheid AX~Cir and its orbiting component.}
  % results heading (mandatory)
   {The companion is resolved by PIONIER at a projected separation $\rho = 29.2 \pm 0.2$\,mas and projection angle $PA = 167.6 \pm 0.3\degr$. We measured $H$-band flux ratios between the companion and the Cepheid of $0.90 \pm 0.10$\,\% and $0.75 \pm 0.17$\,\%, at pulsation phases for the Cepheid of $\phi = 0.24$ and 0.48, respectively. The lower contrast at $\phi = 0.48$ is due to the increased brightness of the Cepheid compared to $\phi = 0.24$. This gives an average apparent magnitude $m\mathrm{_H (comp)} = 9.06 \pm 0.24$\,mag. The limb-darkened angular diameter of the Cepheid at the two pulsation phases was measured to be $\theta_\mathrm{LD} = 0.839 \pm 0.023$\,mas and $\theta_\mathrm{LD} = 0.742 \pm 0.020$\,mas, at $\phi = 0.24$ and 0.48, respectively. A lower limit on the total mass of the system was also derived based on our measured separation, and we found $M_\mathrm{T} \geq 9.7 \pm 0.6 M_\odot$.}
  % conclusions heading (optional), leave it empty if necessary 
   {}

 
 \maketitle

%
%================================================================

\section{Introduction}
\defcitealias{Gallenne_2013_04_0}{Paper I}

Cepheids are powerful astrophysical laboratories that provide fundamental clues for studying the pulsation and evolution of intermediate-mass stars. However, the discrepancy between masses predicted by stellar evolutionary and pulsation models is still not understood well. The most cited scenarios to explain this discrepancy are a mass-loss process during the Cepheid's evolution and/or convective a core overshooting during the main-sequence stage \citep{Neilson_2011_05_0,Keller_2008_04_0,Bono_2006__0}. Therefore, accurate masses of a few percent are needed to help constrain the two models.

So far, the mass of only one Cepheid, Polaris, has been measured \citep{Evans_2008_09_0}; otherwise, they are derived through the mass of the companion inferred from a mass-temperature relation. When in binary systems, Cepheids offer the unique opportunity to make progress in resolving the Cepheid mass problem. The dynamical masses can be estimated \citep{Pietrzynski_2011_12_0, Pietrzynski_2010_11_0, Evans_2008_09_0}, and provide new constraints on evolution and pulsation theory \citep[e.g.][]{Prada-Moroni_2012_04_0}. This gives new insight on the Cepheid mass, and can settle the discrepancy between pulsation and evolution models. Binary systems are also valuable tools to obtain independent distance measurements of Cepheids, needed to calibrate the Leavitt Law.

However, most of the companions are hot main-sequence stars,  and are located too close to the Cepheid ($\sim$1-40 mas) to be observed with a 10-meter class telescope at optical wavelengths. The already existing orbit measurements were estimated only from IUE spectrum or from the radial velocity variations. The only way to spatially resolve such systems is to use long-baseline interferometry or aperture masking. We started a long-term interferometric observing program that aims at studying a sample of northern and southern binary Cepheids. The first goal is to determine the angular separation and the apparent brightness ratio from the interferometric visibility and closure phase measurements. Our long-term objective is to determine the full set of orbital elements, absolute masses and geometric distances. Our program started in 2012 and has already provided new informations on the V1334~Cyg Cepheid system \citep[][hereafter \citetalias{Gallenne_2013_04_0}]{Gallenne_2013_04_0}.

In this second paper, we report the detection of the orbiting companion around the Cepheid \object{AX~Cir} (HD~130701, HR~5527). This pulsating star has a spectroscopic companion, first suspected from composite spectra by \citet{Jaschek_1960_12_0}, and later confirmed by \citet{Lloyd-Evans_1982_06_0}. A preliminary orbital period of about 4600\,days was then estimated by \citet{Szabados_1989_01_0}. \citet{Bohm-Vitense_1985_09_0} and \citet{Evans_1994_11_0} also detected the companion from International Ultraviolet Exporer (IUE) low-resolution spectra, and set its spectral type to be a B6V star. The first orbital solution was provided by \citet{Petterson_2004_05_0} from precise and homogeneous high-resolution spectroscopic measurements; however, it does not include the semi-major axis, the inclination angle, and the longitude of the ascending node, which can only be provided from astrometry. We list some parameters of the AX~Cir system in Table~\ref{table__system_parameters}.

We present here the first spatially resolved detection of this companion from VLTI/PIONIER observations. We first describe in Sect.~\ref{section__observations_and_data_reduction} the beam combiner, the observations, and the raw data calibration. In Sect.~\ref{section__data_analysis} we explain the data analysis and present our results. We then discuss our measured flux ratio and projected separation, and conclude in Sect.~\ref{section__conclusion}.

\begin{table*}[]
\centering
\caption{Parameters of the Cepheid and its close companion.}
\begin{tabular}{ccccccc|ccccccc} 
\hline
\multicolumn{7}{c|}{Primary (Cepheid)}	&	\multicolumn{7}{c}{Secondary\tablefootmark{f}} \\
$\overline{m}_\mathrm{V}$\tablefootmark{a}	&	$\overline{m}_\mathrm{K}$\tablefootmark{b}	&	$\overline{m}_\mathrm{H}$\tablefootmark{b}	&	Sp. Type\tablefootmark{c}	&	$P_\mathrm{pul}$\tablefootmark{c}	&	$\overline{\theta}_\mathrm{LD}$\tablefootmark{d}	&	$d$\tablefootmark{e}	&	Sp. Type	&	$P_\mathrm{orb}$	&	$T_0$ 	&	$e$	&	$a_1\sin{i}$ 	&	$\omega$	&	$f(M)$	\\
	&	&	&	&	(days)	&	(mas)	&	(pc)	&	&	(days)	&	(days)	&	&	(AU)	&	(rad)	&	(M$_\odot$)	\\
\hline
%\noalign{\smallskip}
5.89	&	3.76	&	3.85	&	F8II	&	5.2733	&	0.76	&	500	&	B6V	&	6532	& 2~448~500	&	0.19	& 6.05	&	4.03	&	0.68	\\
\hline
\end{tabular}
\tablefoot{$\overline{m}_\mathrm{V}$, $\overline{m}_\mathrm{K}$, $\overline{m}_\mathrm{H}$: mean apparent $V$, $K$ and $H$ magnitudes. Sp. Type: spectral type. $P_\mathrm{pul}$: period of pulsation. $\overline{\theta}_\mathrm{LD}$: mean angular diameter. $d$: distance. $P_\mathrm{orb}$: orbital period. $T_0$: time passage through periastron. $e$: eccentricity of the orbit. $a_1\ \sin{i}$: projected semi-major axis of the orbit of the Cepheid about the center of mass of the system. $\omega$: argument of periastron. $f(M)$: spectroscopic mass function.\\
\tablefoottext{a}{from \citet{Klagyivik_2009_09_0}.} 
\tablefoottext{b}{from the 2MASS catalog \citep{Cutri_2003_03_0}.} 
\tablefoottext{c}{from \citet{Samus_2009_01_0}.}
\tablefoottext{d}{from  \citet[][at $\phi = 0.27$]{Gallenne_2011_11_0}} 
\tablefoottext{e}{from the $K$-band P-L relation of \citet{Storm_2011_10_0}.} 
\tablefoottext{f}{from \citet{Evans_2000_06_0} and \citet{Petterson_2004_05_0}.}
}
\label{table__system_parameters}
\end{table*}

%================================================================

\section{Observations and data reduction}
\label{section__observations_and_data_reduction}

We used the Very Large Telescope Interferometer \citep[VLTI ;][]{Haguenauer_2010_07_0} with the four-telescope combiner PIONIER \citep{Le-Bouquin_2011_11_0} to measure squared visibilities and closure phases of the AX~Cir binary system. PIONIER combines the light coming from four telescopes in the $H$ band, either in a broad band mode or with a low spectral resolution, where the light is dispersed into three or seven spectral channels. The recombination provides simultaneously six visibilities and four closure phase signals per spectral channel.

\begin{figure}[]
\centering
%\resizebox{\hsize}{!}{\includegraphics{uv.pdf}}
\includegraphics[width = .95\linewidth]{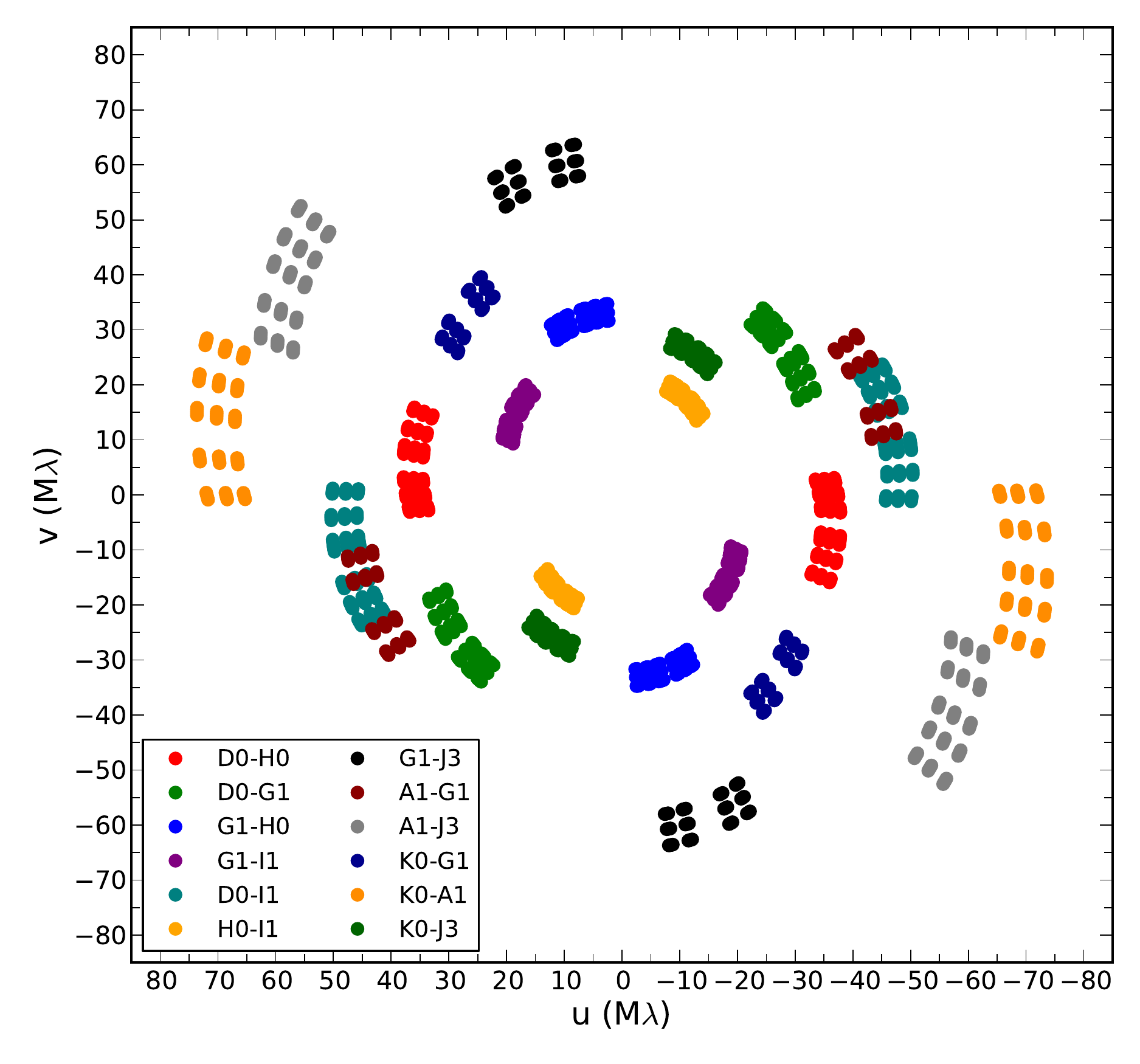}
\caption{$(u,v)$ plane coverage for all our observations of AX~Cir.}
\label{image__uv_plan}
\end{figure}

Our observations were carried out on UT 2013 July 11 and 14, with dispersed fringes in three spectral channels. All observations made use of the 1.8\,m Auxiliary Telescopes with the configuration K0-A1-G1-J3 and  D0-G1-H0-I1, providing six projected baselines ranging from 40 to 140\,m. To monitor the instrumental and atmospheric contributions, the standard procedure, which consists of interleaving the science target by reference stars, was used. The calibrators, HD~133869 and HD~129462,  were selected using the \textit{SearchCal}\footnote{Available at http://www.jmmc.fr/searchcal.} software \citep{Bonneau_2006_09_0,Bonneau_2011_11_0} provided by the JMMC. The journal of the observations is presented in Table~\ref{table__journal} %, and the corresponding calibrators are listed in Table~\ref{table__calibrator}.
and the $(u, v)$ plane covered by the observations is shown in Fig.~\ref{image__uv_plan}. We have collected a total of 435 squared visibility and 300 closure phase measurements.

\begin{figure}[]
\centering
%\resizebox{\hsize}{!}{\includegraphics{CP2.pdf}}
\resizebox{\hsize}{!}{\includegraphics{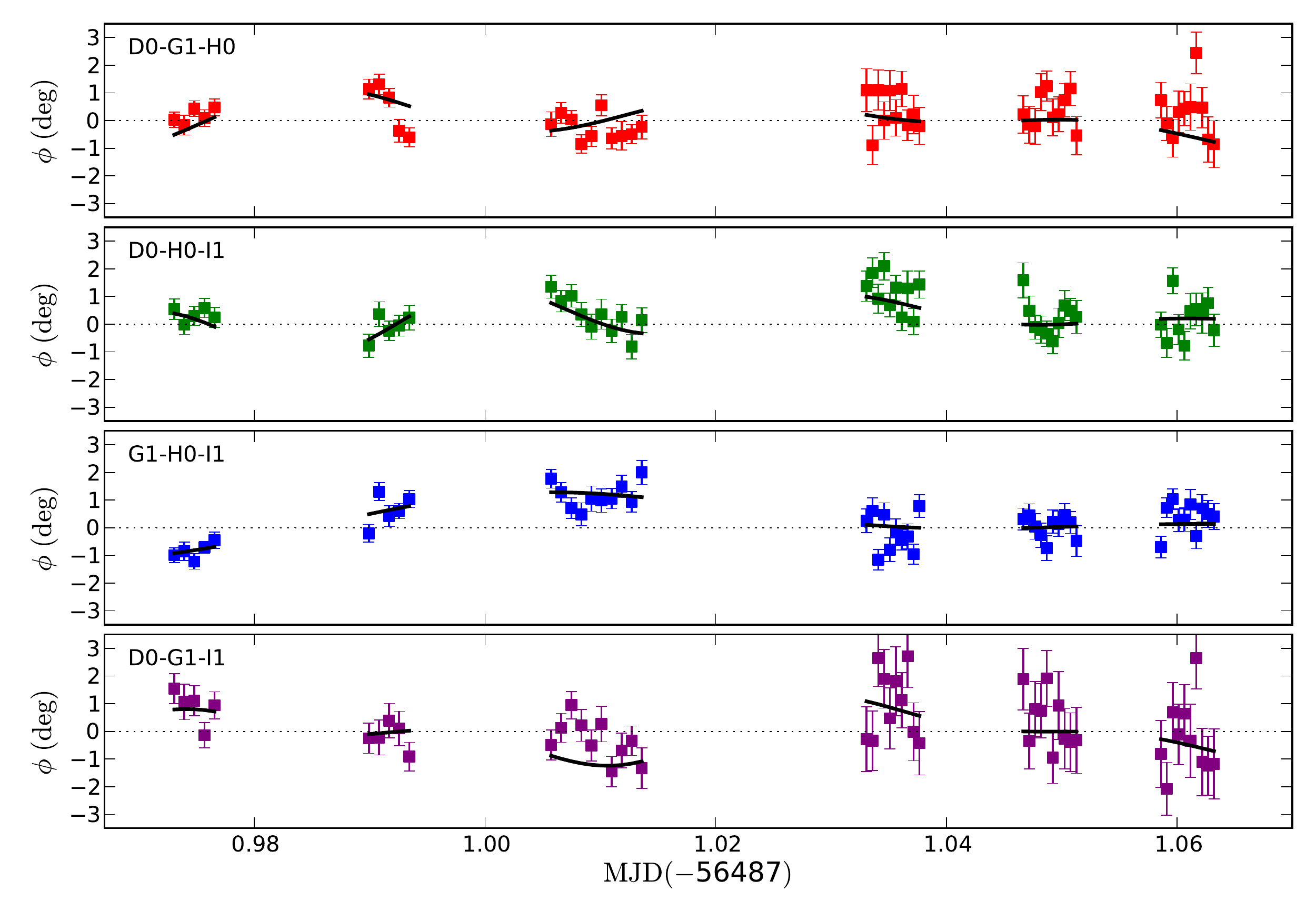}}
\caption{Closure phase signal of AX~Cir for July 14, with respect to the modified Julian date. The spectral channels were averaged for clarity. The solid black line represents our best fit model.}
\label{image__cp}
\end{figure}

The data have been reduced with the \textit{pndrs} package described in \citet{Le-Bouquin_2011_11_0}. The main procedure is to compute squared visibilities and triple products for each baseline and spectral channel, and to correct for photon and readout noises. The final calibrated closure phases of July 14 are presented in Fig~\ref{image__cp}. The variations in the signal suggest the presence of the companion, and this is strengthened by a higher signal-to-noise ratio when combining all the data.

\onltab{
\begin{table}[]
\centering
\caption{Journal of the observations.}
\begin{tabular}{cccc} 
\hline
\hline
%UT 				&													&	Star				&		Configuration			\\
 \multicolumn{2}{c}{UT} 															&	Star				&		Configuration			\\
\hline
2013~July~11	&	0:14 & AX~Cir 	&   K0-A1-G1-J3 		\\
							&	0:27 & HD~129462 	&   K0-A1-G1-J3 		\\
							&	0:36 & AX~Cir 	&   K0-A1-G1-J3 		\\
							&	0:46 & HD~129462 	&   K0-A1-G1-J3 		\\
							&	0:58 & AX~Cir 	&   K0-A1-G1-J3 		\\
							&	1:18 & HD~133869 	&   K0-A1-G1-J3 		\\
							&	1:27 & AX~Cir 	&   K0-A1-G1-J3 		\\
							&	1:39 & HD~133869 	&   K0-A1-G1-J3 		\\
							&	1:51 & AX~Cir 	&   K0-A1-G1-J3 		\\
							&	2:01 & HD~133869 &   K0-A1-G1-J3 		\\
2013~July~14	&	23:07 & \object{HD~133869} 	&   D0-G1-H0-I1 		\\ 
							 & 23:21 &	AX~Cir	&	  D0-G1-H0-I1 		\\
							 & 23:32 &	\object{HD~129462}  &	  D0-G1-H0-I1 		\\
							 & 23:45 &	AX~Cir		 		&	  D0-G1-H0-I1 		\\
							 & 23:55 &	HD~133869  &	  D0-G1-H0-I1 		\\
2013~July~15 	& 00:08 &	AX~Cir		 		&	  D0-G1-H0-I1 		\\
							 & 00:24 &	HD~129462  &	  D0-G1-H0-I1 		\\
							& 00:47 &	AX~Cir		 		&	  D0-G1-H0-I1 		\\
							 & 00:58 &	HD~133869  &	  D0-G1-H0-I1 		\\
							& 01:07 &	AX~Cir		 		&	  D0-G1-H0-I1 		\\
							 & 01:17 &	HD~129462  &	  D0-G1-H0-I1 		\\
							& 01:24 &	AX~Cir		 		&	  D0-G1-H0-I1 		\\
							 & 01:35 &	HD~129462  &	  D0-G1-H0-I1 		\\
\hline
\end{tabular}
\label{table__journal}
\end{table}
}

%\begin{table}[]
%\centering
%\caption{Calibrators used for our observations.}
%\begin{tabular}{cccccc} 
%\hline
%\hline
%Calibrator  &	$m_\mathrm{V}$	& 	$m_\mathrm{H}$	&	Sp.~Type	& 	$\theta_\mathrm{UD}$	&	$\gamma$		\\
%(HD)		&								&							&						&	(mas)				&	($\degr$)	\\
%\hline		
%%\multicolumn{6}{c}{V1334~Cyg} \\
%132905		& 	5.2  					&  3.1				&  G8III  	&  $1.184 \pm 0.084$ 		& 1.4			\\ 
%133869		& 	8.0  					&  3.7				&  K3III  	&  $1.043 \pm 0.015$ 		& 3.0			\\ 
%129462		& 	6.1  					&  3.7				&  K0III  	&  $0.857 \pm 0.061$ 		& 5.4			\\ 
%\hline
%\end{tabular}
%\tablefoot{$m_\mathrm{V}, m_\mathrm{H}$: magnitudes in $V$ and $H$ bands. $\theta_\mathrm{UD}$: uniform disk angular diameter in $H$ band. $\gamma$: angular distance to the Cepheid.\\
%}
%\label{table__calibrator}
%\end{table}

\begin{figure}[!h]
\centering
\resizebox{\hsize}{!}{\includegraphics{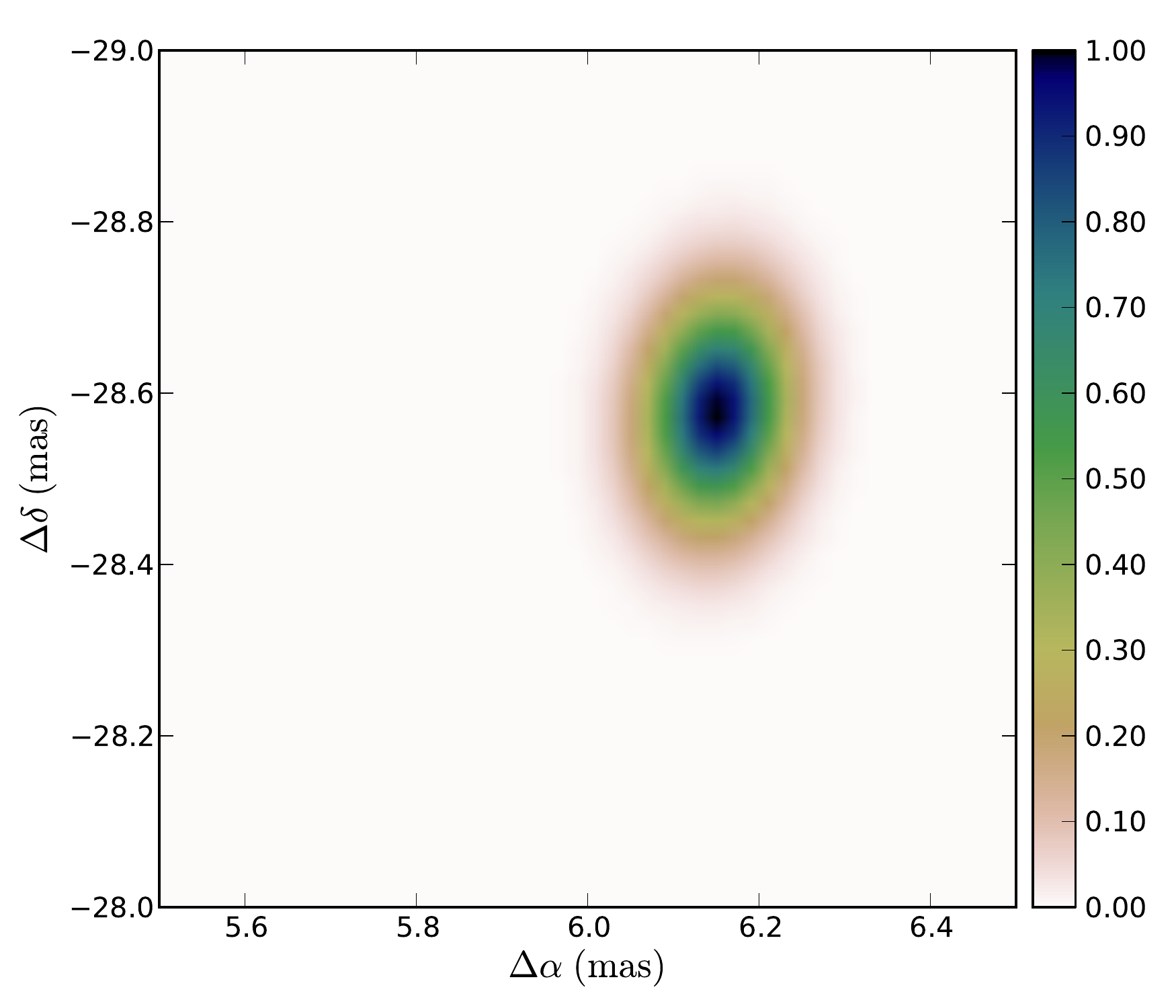}}
\caption{Probability map for the companion position of July 14.}
\label{image__chi2_map}
\end{figure}

\begin{figure}[]
\centering
\resizebox{\hsize}{!}{\includegraphics{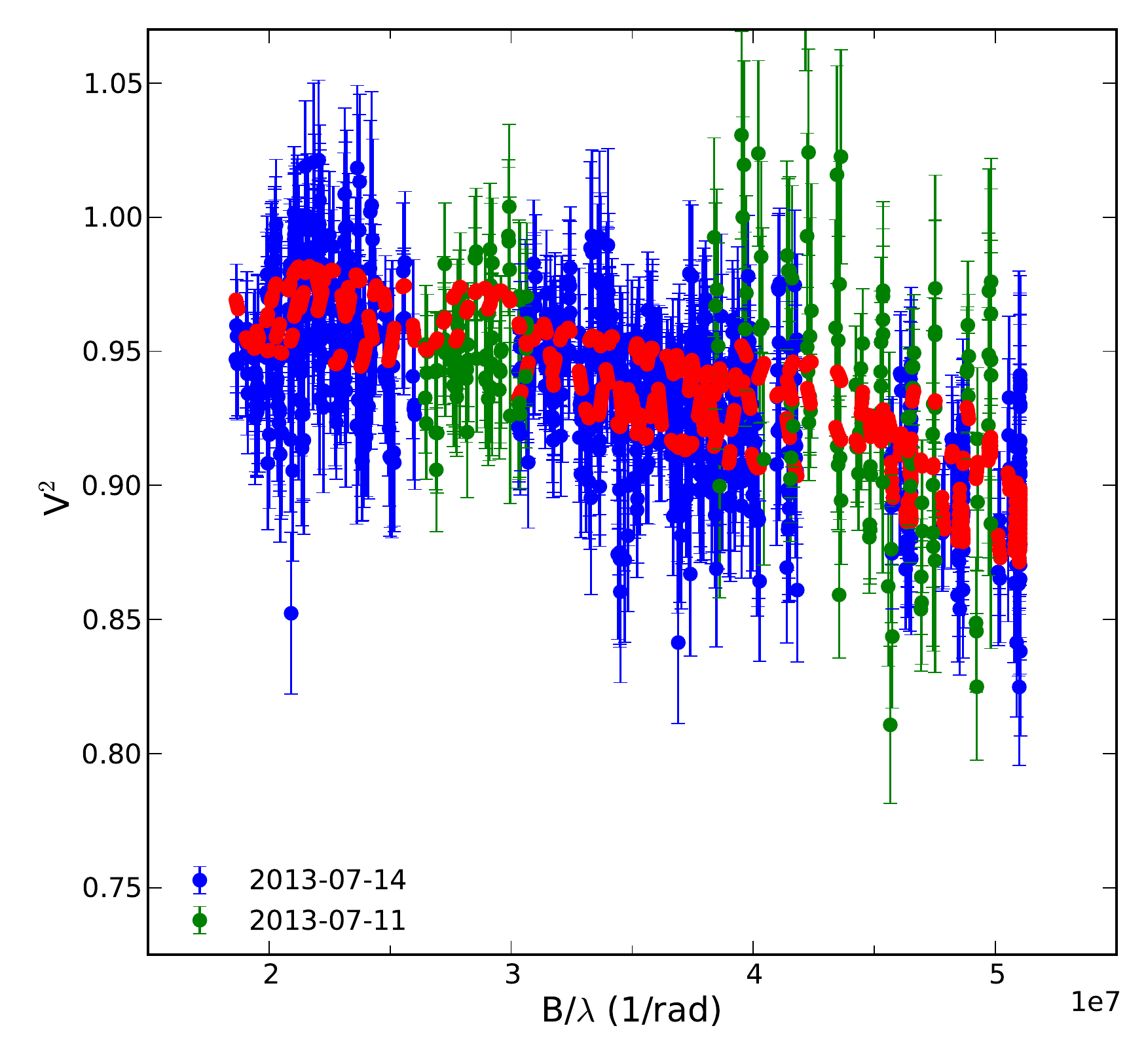}}
\caption{Squared visibility measurements of AX~Cir. The data are in blue for July 14 and in green for July 11, while the red dots are the fitted binary model for both epochs.}
\label{image__visibility}
\end{figure}

%================================================================

\section{Model fitting}
\label{section__data_analysis}

%To model the squared visibilities and closure phase signals, we used the \textit{LITpro}\footnote{LITpro software available at http://www.jmmc.fr/litpro.} model fitting software \citep{Tallon-Bosc_2008_07_0}, based on the Levenberg-Marquardt algorithm. It provides a set of elementary models that can be combined all together. The software also contains a tool allowing the search for the global minimum to solve for the problem of multiple $\chi^2$ minima.

The squared visibilities and closure phase signals were modeled assuming a uniform disk (UD) angular diameter for the Cepheid (the primary) plus a point source companion. The fitted parameters are the angular diameter of the Cepheid $\theta_\mathrm{UD}$, the relative position of the component ($\Delta \alpha$, $\Delta \delta$), and the flux ratio $f = f_\mathrm{com}/f_\mathrm{cep}$. The coherence loss effect due to spectral smearing on the companion was also modeled using the function $|\mathrm{sinc}x|$, where $x = \pi (u \Delta \alpha + v \Delta \delta) / (\lambda R)$ at spectral resolution $R = 18$ and spatial frequencies $(u, v)$.

The choice of a UD diameter for the Cepheid instead of a limb-darkened (LD) disk for the fitting procedure is justified because the angular diameter is small compared to the angular resolution of the interferometer, and the limb darkening effects are therefore undetectable. The conversion from UD to LD angular diameter was done afterwards by using a linear-law parametrization $I_\lambda (\mu) = 1 - u_\lambda(1 - \mu)$, with the LD coefficient $u_\lambda = 0.2887$ \citep{Claret_2011_05_0} for both epochs, and using the stellar parameters $T_\mathrm{eff} = 5400$\,K, $\log g = 2.0$, [Fe/H] = 0.0, and $v_\mathrm{t} = 5$\,km~s$^\mathrm{-1}$ \citep{Usenko_2011_07_0,Acharova_2012_02_0}. The conversion is then given by the approximate formula of \citet{Hanbury-Brown_1974_06_0}:
\begin{displaymath}
\theta_\mathrm{LD}(\lambda) = \theta_\mathrm{UD}(\lambda) \sqrt{\frac{1-u_\lambda/3}{1-7u_\lambda/15}}.
\end{displaymath}

Changing $T_\mathrm{eff}$ by $\pm 400$\,K changes the diameter by less than 0.2\,\%, well below our measured uncertainties.

For each epoch, the fitting procedure was done in two steps. We first proceeded to a 80x80\,mas grid search in the $\chi^2$ space, with spacing of 0.2\,mas, which aims at determining the approximate position of the companion and avoid local minima. Then a finer search of 5x5\,mas with a 0.05\,mas spacing around the most likely position was carried out to obtain the final parameters. We chose $\theta_\mathrm{UD} = 0.76$\,mas \citep{Gallenne_2011_11_0} and $f = 1.5$\,\% \citep{Evans_1994_11_0} as first guesses.

Our model did not take a possible circumstellar envelope (CSE) emission into account, which could lead to an overestimate of the angular diameter. From the spectral energy distribution AX~Cir given by \citet{Gallenne_2011_11_0}, the infrared excess caused by the CSE appears around $10\,\mathrm{\mu m}$, while it is negligible at $1.6\,\mathrm{\mu m}$ (i.e., $< 2$\,\%, which would lead to visibility loss of the same amount at first order, and below our visibility accuracy).

%The closure phase signal presented in Figs.~\ref{image__cp} clearly show a departure from a single star with a symmetric brightness distribution. This is a signature of an orbiting companion.

The probability map for the observations of July 14 is shown in Fig.~\ref{image__chi2_map}, and the fitted parameters for both epochs are reported in Table~\ref{table__fitted_parameters}. The companion is clearly detected at the two epochs at coordinates $\rho = 29.2 \pm 0.2$\,mas and $PA = 167.6 \pm 0.3\degr$. The model for the observations of July 14 is represented graphically in Fig~\ref{image__cp}. We estimated limb-darkened angular diameters $\theta_\mathrm{LD} = 0.742 \pm 0.020$\,mas and $0.839 \pm 0.023$\,mas, for July 11 and 14, respectively (at pulsation phases $\phi = 0..48$ and 0.24, respectively), in agreement with the angular diameter, $0.76 \pm 0.03$ estimated by \citet{Gallenne_2011_11_0} at phase $\phi = 0.27$. It is also consistent with the average value of 0.84\,mas estimated from the surface brightness relation of  \citet[][using magnitudes from Table~\ref{table__system_parameters}]{Kervella_2004_12_0}. However, no IR photometric measurements were available at the time of our interferometric observations, and we cannot compare our measured diameters to those derived from surface brightness relationships. Uncertainties were estimated using the subsample bootstrap technique with replacement and 10~000 subsamples. The medians of the probability distribution of the parameters match the best-fit values very well, and we used the maximum value between the 16\,\% and 84\,\% percentiles as uncertainty estimates (although the distributions were roughly symmetrical about the median values). %\textbf{We also added quadratically the systematics due to the errors in calibrators size.}

\begin{table}[]
\centering
\caption{Final best-fit parameters.}
\begin{tabular}{ccc} 
\hline
\hline
														&	2013-07-11							&	2013-07-14\\
\hline
Single star model 							&												&	  \\
$\theta_\mathrm{UD}$ (mas)		&	$0.770 \pm 0.016$   	 &	$0.931 \pm 0.019$         	\\
$\theta_\mathrm{LD}$ (mas)			&  $0.787 \pm 0.016$  	 &  $0.952 \pm 0.020$  		          	\\
$\chi^2_r$										&	   1.45								&	1.09            								\\
\hline
Binary model \\
$\theta_\mathrm{UD}$ (mas)				&	$0.726 \pm 0.020$  	 &	$0.821 \pm 0.022$          	\\
$\theta_\mathrm{LD}$ (mas)			&	$0.742 \pm 0.020$  		 &	$0.839 \pm 0.023$  	         	\\
$f$ (\%)											&	$0.75 \pm 0.17$				&	$0.90 \pm 0.10$      		\\
$\Delta \alpha$	(mas)						&	$6.421 \pm 0.198$     &	$6.153 \pm 0.155$     	 	\\
$\Delta \delta$	(mas)				  	 &	$-28.366 \pm 0.366$  	   	&	$-28.584  \pm 0.229$  	   	\\
$\chi^2_r$											&	   1.17       						&	0.72  									\\
\hline
\end{tabular}
\tablefoot{$\theta_\mathrm{UD}$, $\theta_\mathrm{LD}$: uniform and limb-darkened disk angular diameter, respectively. $f$, $\Delta x$, $\Delta y$: flux ratio and position of the companion. $\chi^2_r$: reduced $\chi^2$ of the corresponding best-fit model.}
\label{table__fitted_parameters}
\end{table}

%================================================================

\section{Discussion}

The measured flux ratios are slightly different between the two epochs, although within the uncertainties. This is because the Cepheid is slightly brighter at phase $\phi = 0.48$ (July 11) than in $\phi = 0.24$, which makes the contrast a bit lower. Since we do not have $H$-band light curves to extract the Cepheid magnitude at a given phase, we took an average to estimate a mean contrast $f = 0.83 \pm 0.14$\,\%. This gives a difference in apparent magnitude of $\Delta m_\mathrm{H} = 5.20 \pm 0.18$\,mag. This converts to apparent magnitudes for each component by using the 2MASS magnitude as a measure of the combined flux and the following equations:
\begin{eqnarray}
m_1 &=& m_{12} + 2.5\log(1 + f)\\
m_2 &=& m_{12} + 2.5\log(1 + 1/f)
\end{eqnarray}
where $m_{12}$ is the 2MASS measurements, and $m_1$ and $m_2$ the apparent magnitude of the Cepheid and the component, respectively.  We obtain $H_\mathrm{comp} = 9.06 \pm 0.24$\,mag and $H_\mathrm{cep} = 3.86 \pm 0.24$\,mag. The quoted errors are due to the uncertainties in 2MASS.  We determined the dereddened magnitude, $H_{0}^\mathrm{comp} = 8.94 \pm 0.24$\,mag and $H_{0}^\mathrm{cep} = 3.72 \pm 0.24$\,mag, by adopting the reddening law from \citet{Fouque_2007_12_0} with a total-to-selective absorption in the $V$ band of $R_\mathrm{V} = 3.23$ \citep{Sandage_2004_09_0} and a color excess $E(B - V) = 0.262$ from \citet{Tammann_2003_06_0}. From the distance $d = 500 \pm 10$\,pc given by the $K$-band period--luminosity relation \citep[][the quoted error is statistical]{Storm_2011_10_0}, we obtain an absolute magnitude for the companion $M_\mathrm{H} = 0.45 \pm 0.24$\,mag. Combining the known spectral type B6V with a color-spectral type relation \citep{Ducati_2001_09_0}, we obtain $M_{V} = -0.12 \pm 0.24$\,mag. %This value is $2\sigma$ away from the the 6.92\,mag estimated by \citet[][converted from UV observations]{Evans_1994_11_0}. This discrepancy might come from the magnitude conversion from $H$ to $V$ (this work) or from $UV$ to $V$ \citep{Evans_1994_11_0}.

%A range of inclination angles can be set assuming our measured projected separation as a lower limit for the angular semi-major axis. We use the following equation:
%\begin{displaymath}
%\rho = a\,(1 - e\cos E) \sqrt{1 - \sin^2i\,\sin^2(\omega + \nu)},
%\end{displaymath}
%with $\rho$ the measured angular separation, $a$ angular semi-major axis, $e$ the eccentricity, $E$ the eccentric anomaly, $i$ the inclination angle, $\omega$ the argument of periastron, and $\nu$ the true anomaly. The condition $a \geq \rho$ leads to:
%\begin{displaymath}
%| \sin i | \geq \frac{\sqrt{1 - (1 - e\cos E)^{-2}}}{|\sin(\omega + \nu)|}
%\end{displaymath}
%
%Taking the spectroscopic orbital elements from \citet{Petterson_2004_05_0} and restricting to values $0 < i < \pi$, we found that $ i \geq 11.1 \pm 0.1^\circ$. 

From Kepler's law and  assuming our measured projected separation $\rho$ as a lower limit for the angular semi-major axis, that is $a \geq \rho$, a minimal total mass for the system can be derived: 
\begin{displaymath}
M_\mathrm{T} = M_\mathrm{1} + M_\mathrm{2} \geq \frac{\rho^3 d^3}{P^2},
\end{displaymath}
with $\rho$ in arcsecond, $d$ in parsec, and $P$ in year. We therefore derived $M_\mathrm{T} \geq 9.7 \pm 0.6\,M_\odot$. This is compatible with the $5.1\,M_\odot$ for the Cepheid, predicted from the pulsation mass \citep{Caputo_2005_08_0}, and with the $5\,M_\odot$ for the companion, inferred from its spectral type.

%Using the distance $d = 500 \pm 10$\,pc given by the $K$-band Period--Luminosity relation \citep{Storm_2011_10_0}, we obtain an absolute magnitude for the companion $M_\mathrm{H} = 0.34 \pm 0.24$\,mag. Combining the known spectral type B6V and the color-spectral type relation derived by \citet{Ducati_2001_09_0}, we have  in the visible $M_\mathrm{V} = -0.24 \pm 0.24$\,mag. This is not consistent with  $M_{V} = -1.96$\,mag derived by \citet{Bohm-Vitense_1985_09_0} and the one of \citet{Evans_1994_11_0}, however this strongly depends on the adopted reddening law and distance. At that time they used a distance of about 360\,pc, well smaller than our recent value of 500\,pc.

%The companion is a B6V star \citep{Evans_1994_11_0}

%================================================================

\section{Conclusion}
\label{section__conclusion}

We used the high angular resolution provided by the four-telescope combiner PIONIER to detect the orbiting companion of the short-period cepheid AX~Cir. We employed a binary model with a primary represented by a uniform disk and the secondary as an unresolved source. We derived a limb-darkened angular diameter for the Cepheid at two pulsation phases, $\theta_\mathrm{LD} = 0.839 \pm 0.023$\,mas (at $\phi = 0.24$) and $\theta_\mathrm{LD} = 0.742 \pm 0.020$\,mas (at $\phi = 0.48$). We also measured an averaged $H$-band flux ratio between the companion and the Cepheid, $f = 0.83 \pm 0.14$\,\%, and the astrometric position of the secondary relative to the primary, $\rho = 29.2 \pm 0.2$\,mas and $PA = 167.8 \pm 0.3\degr$. We also set a lower limit on the total mass of the system based on our measured projected separation. Finally, we point out the need of accurate infrared light curves to enable a more precise flux estimate of the companion from the contrast measured from interferometry.

This second detection (after that of V1334~Cyg, \citetalias{Gallenne_2013_04_0}) demonstrates the capabilities of long-baseline interferometers for studying the close-orbit companions of Cepheids. Further interferometric observations will be obtained in the future to cover the orbit, and then combined with radial velocity measurements to derive all orbital elements. For now, only single-line spectroscopic measurements are available, we are also involved in a long-term spectroscopic program to detect the radial velocity of the companion. This will provide an orbital parallax and model-free masses.

%--------------------ACKNOWLEDGEMENTS--------------------

\begin{acknowledgements}
The authors thank all the people involved in the VLTI project. AG acknowledges support from FONDECYT grant 3130361. JDM acknowledges funding from the NSF grants AST-1108963 and AST-0807577. WG and GP gratefully acknowledge financial support for this work from the BASAL Centro de Astrof\'isica y Tecnolog\'ias Afines (CATA) PFB-06/2007. Support from the Polish National Science Center grant MAESTRO and the Polish Ministry of Science grant Ideas Plus (awarded to GP) is also acknowledged. We acknowledge financial support from the “Programme National de Physique Stellaire” (PNPS) of CNRS/INSU, France. PIONIER was originally funded by the Poles TUNES and SMING of Universit\'e Joseph Fourier (Grenoble) and subsequently supported by INSU-PNP and INSU-PNPS. The integrated optics beam combiner is the result of collaboration between IPAG and CEA-LETI based on CNES R\&T funding. This research received the support of PHASE, the high angular resolution partnership between ONERA, the Observatoire de Paris, CNRS, and University Denis Diderot Paris 7. This work made use of the SIMBAD and VIZIER astrophysical database from the CDS, Strasbourg, France, and the bibliographic informations from the NASA Astrophysics Data System. This research has made use of the Jean-Marie Mariotti Center \texttt{SearchCal} service, co-developed by FIZEAU and LAOG/IPAG. The research leading to these results received funding from the European Research Council under the European Community's Seventh Framework Program (FP7/2007--2013)/ERC grant agreement n$^\circ$227224 (PROSPERITY).
\end{acknowledgements}

%--------------------BIBLIOGRAPHY--------------------

\bibliographystyle{aa}   % if natbib is available
\bibliography{/Users/alex/Sciences/Articles/bibliographie}

\end{document}